\documentclass[preprint,showpacs,nofootinbib,prd,aps]{revtex4-1}
\usepackage{graphicx,amstext,amssymb,amsmath,color}
\begin{document}

\title{Quasi equilibrium state of expanding  quantum fields  and  two-pion Bose-Einstein correlations
 in $pp$ collisions at the LHC}

\author{S.V. Akkelin$^{1}$}
\affiliation{$^1$Bogolyubov Institute for Theoretical Physics,
Metrolohichna  14b, 03143 Kiev,  Ukraine}

\begin{abstract}

We argue that  the two-particle momentum correlation functions of
high-multiplicity $p+p$ collisions at the LHC provide a signal for a
ground state structure of a quasi equilibrium state of the
longitudinally boost-invariant expanding quantum field which lies in
the future light cone of a  collision. The physical picture   is
that pions are produced by the  expanding quantum emitter with two
different  scales approximately attributed to the expanding ideal
gas in local equilibrium state  and ground-state condensate.
Specifically, we show that the effect of suppressing the
two-particle Bose-Einstein momentum correlation functions increases
with increasing transverse momentum of a like-sign pion pair due to
different momentum-dependence of  the corresponding particle
emission regions.

\end{abstract}


 \maketitle

 \section{Introduction}

It is firmly established now that collective phenomena in
relativistic heavy ion collisions are associated with hydrodynamics
\cite{Hydro-1-1,Hydro-1-2,Hydro-1-3}. Surprisingly, similar
collective phenomena have been observed recently in
high-multiplicity $p+p$ collisions at the CERN Large Hadron Collider
(LHC) \cite{Hydro-LHC}. It is not clear however whether such
collective phenomena can be attributed to hydrodynamic evolution
like in $A+A$ collisions \cite{Hydro-2-1,Hydro-2-2}. Also, while
there is some evidence that hydrodynamics can be successfully
applied to describe flow-like features in high-multiplicity $p+p$
collisions, see e.g. refs.
\cite{Hydro-pp-1-1,Hydro-pp-1-2,Hydro-pp-2-1,Hydro-pp-2-2} (for
discussions on whether hydrodynamics is applicable in small systems
see e.g. refs. \cite{Rom-1,Rom-2}), till now there is no united
description of one-particle momentum spectra  and multi-particle
momentum correlations in $p+p$ collisions in any detail dynamical
model.

Many-particle correlations  play an important role in the
understanding of multiparticle production mechanisms. In particular,
the correlation femtoscopy method (commonly referred to as
femtoscopy, or HBT interferometry) uses momentum correlations of two
identical particles at low relative momenta to extract information
about the space-time evolution  and properties of the expanding
matter in high energy nucleon and nuclear collisions, for reviews
see e.g. refs. \cite{HBT-1,HBT-2,HBT-3,HBT-4,HBT-5,HBT-6,HBT-7}.
Because in such collisions most of produced particles are pions, the
Bose-Einstein correlations of two identical pions are usually
analyzed. The measured Bose-Einstein correlations of particles with
some fixed momentum of a particle pair are  typically described by a
function with two sort of parameters: the effective radius
parameters $R_{i}$ (sometimes called ``HBT radii''), which can be
interpreted as mean (statistically averaged) distance between
centers of particle emissions, and the ``correlation strength''
parameter $\lambda$, the latter is also called the incoherence or
chaoticity parameter. The interpretation of the latter  is not
unambiguous, because under the real experimental conditions
$\lambda$-parameter can be affected by particle misidentifications,
decays of long-lived resonances\footnote{ Loosely speaking, then
typical interferometry radii of a detected pair are very large and
cannot be resolved because the resolution requires very low
difference of momenta of detected particles beyond the experimental
limits, in a result, corresponding quantum statistical momentum
correlations of particles are not seen.}, and non-Gaussian features
of the particle distributions. Practically, the values of
$\lambda$-parameter obtained fitting experimental data are
momentum-dependent and less than $1$.

Results from femtoscopic studies of two-particle  correlations from
$p+p$ collisions at the LHC have been presented in refs.
\cite{Alice-1,Atlas,CMS}. It was found that the femtoscopic radii
measured in high-multiplicity $p+p$ collisions are smaller than the
ones in relativistic heavy ion collisions, and decrease with
increasing momentum of a pair. The latter in hydrodynamical approach
can be interpreted as the decrease of ``lengths of homogeneity''
\cite{Sin-2-1,Sin-2-2,Sin-2-3,Sin-2-4} (sizes of the effective
emission region). This is a direct consequence of generated by
hydrodynamical flow $x-p$ correlations: Due to the collective flow
pion pairs of higher momenta are effectively emitted from  smaller
regions of the source. Also, it was found that $\lambda$-parameter
is essentially less than unity, and exhibits a decrease as the
momentum of a like-sign pion pair increases \cite{Atlas,CMS}. This
seems to be at variance with the behavior of pionic $\lambda$
parameter in reliable hydrodynamical models of $A+A$ collisions, see
e.g. ref. \cite{Shapoval}.

Hydrodynamic is, in essence, the  local conservation equations of
the expectation values of the stress-energy tensor and charge
current operators. In particular, at the lowest order in the
gradient expansion the mean value of the stress-energy tensor takes
the ideal fluid form. It is worth noting  that hydrodynamic
equations can be derived using Zubarev's formalism of
non-equilibrium statistical operator
\cite{Zubarev-1,Zubarev-2,Zubarev-3,Zubarev-4,Zubarev-5}, see also
refs. \cite{Bec-1,Bec-2,Bec-3,Bec-4} for recent papers related to
this method. The corresponding statistical operator is appropriate
for reduced description of the system and can be used to calculate
expectation values of relevant observables. Even if the statistical
operator can be approximated by its quasi equilibrium form for
non-interacting quantum fields, the corresponding expressions can
demonstrate essential deviations from the ones at the global
thermodynamic equilibrium with a temperature, a collective
four-velocity, and a chemical potential equal to those in some
space-time point, see refs.
\cite{Sinyukov-1,Sinyukov-1-1,Sinyukov-2}. In the present paper,
using a simple and physically clear method of calculations we
re-calculate particle momentum spectra and Bose-Einstein
correlations  first evaluated for a boost-invariant expanding free
boson quantum field in refs.
\cite{Sinyukov-1,Sinyukov-1-1,Sinyukov-2}. It allows us, in
particular, to reveal the underlying physical mechanism which is
responsible for the non-trivial quasi equilibrium ground state of
expanding non-interacting quantum fields.  Then, unlike  refs.
\cite{Sinyukov-1,Sinyukov-1-1,Sinyukov-2} where ground-state
condensate contributions of quasi equilibrium statistical operator
to particle momentum spectra were treated as physically meaningless
and therefore subtracted from the particle spectra, in the present
paper we argue its physical significance and possible relevance to
the observed effect of suppression of the two-particle Bose-Einstein
momentum correlation functions  in high-multiplicity $p+p$
collisions.

\section{Quasi equilibrium state  of  boost-invariant expanding quantum  scalar field}

Following refs.
\cite{Zubarev-1,Zubarev-2,Zubarev-3,Zubarev-4,Zubarev-5}, we  define
a quasi equilibrium  statistical operator on Minkowski spacetime  as
\begin{eqnarray}
\rho=Z^{-1} \exp\left(-\int_{\sigma_{\nu}} d\sigma n_{\nu}(x)\beta
(x)u_{\mu}(x)T^{\mu \nu}(x)\right ), \label{1}
\end{eqnarray}
where $\beta (x)= 1/T(x)$ is the local inverse temperature,
$u_{\mu}(x)$ is hydrodynamical four-velocity,
$u_{\mu}(x)u^{\mu}(x)=1$,\footnote{For the Minkowskian metric tensor
 we use the convention $g^{\mu \nu}=\mbox{diag}(+1,-1,-1,-1)$.}
$\sigma_{\nu}$ is a three-dimensional space-like hypersurface with a
time-like normal vector $n_{\nu}(x)$, $T^{\mu \nu}(x)$ is the
operator of an energy-momentum tensor, and $Z$ is the normalization
factor making $Tr[\rho]=1$. We assume for simplicity that a chemical
potential $\mu = 0$. The expectation value of an operator
$\widehat{O}$ can be expressed as
\begin{eqnarray}
\langle  \widehat{O }\rangle = Tr[\rho  \widehat{O}]. \label{2}
\end{eqnarray}
The quasi equilibrium statistical operator  $\rho$ is determined
from the maximization of the information entropy with specific
constraints on the average local values of the energy-momentum
density  operator on the hypersurface $\sigma_{\nu}$: $n_{\mu} (x)
\langle T^{\mu \nu}(x)\rangle$, see e.g. refs.
\cite{Zubarev-1,Zubarev-2,Zubarev-3,Zubarev-4,Zubarev-5,Bec-1,Bec-2,Bec-3,Bec-4}.
Notice that because field operators commute for space-like
separations, the corresponding operators are clearly local
observables.

To apply this formalism to high-multiplicity $p+p$ scatterings at
the LHC, we take into account that hydrodynamics with
boost-invariance in the longitudinal direction \cite{Bjorken} give
reliable results for one-particle momentum spectra in $p+p$
collisions \cite{Hydro-pp-2-1,Hydro-pp-2-2}.  Also, aiming to
perform calculations analytically in some tractable approximations,
we neglect finite-size effects in transverse directions (collective
expansion, inhomogeneities, etc.). Then, assuming  that matter
produced in a high energy $p+p$ collision is locally restricted to
the light cone with beginning at $t=z=0$ plane of the Minkowski
spacetime manifold, we utilize  Bjorken coordinates $(\tau , \eta)$
instead of Cartesian ones, $(t,z)$. Namely, we define
\begin{eqnarray}
t= \tau \cosh \eta, \label{3} \\
z= \tau \sinh  \eta .  \label{4}
\end{eqnarray}
The two other coordinates $\textbf{r}_{T}=(r_{x},r_{y})$ are the
Cartesian ones.  The Minkowski line element restricted to the light
cone is
\begin{eqnarray}
ds^{2}=dt^{2}-d\textbf{r}_{T}^{2}-dz^{2}=  d\tau^{2}- \tau^{2}d
\eta^{2} - d\textbf{r}_{T}^{2}. \label{5}
\end{eqnarray}
Such a coordinate system is nothing but the Milne frame \cite{Milne}
and is associated  with the system of the (hypothetical) observers
which move with different but constant longitudinal velocities in
such a way that their world lines begin at $z=t=0$. Then
boost-invariant hydrodynamic four-velocity $u_{\mu}$ reads
\begin{eqnarray}
u^{\mu}(x)=(\cosh \eta , 0,0, \sinh  \eta). \label{5.1}
\end{eqnarray}
Because small systems created in $p+p$ collisions do not exhibit
prolonged post-hydrodynamical kinetic stage of hadronic
rescatterings, we assume that particle momentum spectra are frozen
immediately after emission (so called sharp freeze-out assumption
\cite{Cooper}) at a hypersurface with constant energy density in the
comoving coordinate system. Then $\beta (x)$ is constant on the
corresponding hypersurface. Such a hypersurface is defined by
constant $\tau = \sqrt{t^{2}-z^{2}}$ \cite{Bjorken}, that is proper
time of the (hypothetical) inertial local observers comoving with a
fluid element with a constant rapidity $\eta$ and constant
transverse coordinates $\textbf{r}_{T}$. This implies that
\begin{eqnarray}
n^{\mu}(x)=u^{\mu}(x), \label{6}
\end{eqnarray}
and
\begin{eqnarray}
d \sigma  = \tau d\eta dr_{x}dr_{y}. \label{7}
\end{eqnarray}
It is worth noting that $\tau$ also controls a value of the
collective velocity space-time gradients,
\begin{eqnarray}
\partial_{\mu}u^{\mu}=\frac{1}{\tau}. \label{7.1}
\end{eqnarray}

To proceed further, one needs to specify  an energy-momentum tensor
in eq. (\ref{1}). Because main aim  of this paper is to reveal a
ground-state condensate  associated with quasi equilibrium expansion
of quantum fields, and then study its effects on  the
single-particle momentum spectra and the two-particle momentum
correlation functions, we neglect here field self-interactions and
use  simple scalar quantum field model with the classical action
\begin{eqnarray}
S=\int dt d^{3}r \left [\frac{1}{2} \left (\frac{\partial
\phi}{\partial t}\right )^{2} - \frac{1}{2} \left (\frac{\partial
\phi}{\partial \textbf{r}}\right )^{2} - \frac{m^{2}}{2}\phi^{2}
\right ] \equiv \int dt d^{3}r L, \label{8}
\end{eqnarray}
where $L$ is the corresponding Lagrangian density.  Then
\begin{eqnarray}
T^{\mu \nu}(x)=\partial^{\mu}\phi\partial^{\nu}\phi -g^{\mu \nu}L,
\label{9}
\end{eqnarray}
and, using eqs. (\ref{3}), (\ref{4}), (\ref{5.1}), (\ref{6}),
(\ref{7.1}), and (\ref{9})  we get
\begin{eqnarray}
n_{\mu}u_{\nu} T^{\mu\nu}(x)= u_{\mu}u_{\nu} T^{\mu\nu}(x)= \left
(\frac{\partial\phi}{\partial\tau}\right )^{2} - L, \label{9.1}
\end{eqnarray}
where we took  into account that
$u_{\mu}\partial^{\mu}=\partial_{\tau}$. The Lagrangian density $L$
in the  Bjorken coordinates (then $dt d^{3}r = \tau d \tau d \eta
d^{2}r_{T} $) is
\begin{eqnarray}
 L = \frac{1}{2} \left (\frac{\partial \phi}{\partial \tau}\right
 )^{2} -
\frac{1}{2} \frac{1}{\tau^{2}}\left (\frac{\partial \phi}{\partial
\eta}\right )^{2} -  \frac{1}{2}\left (\frac{\partial \phi}{\partial
\textbf{r}_{T}}\right )^{2} - \frac{1}{2}m^{2}\phi^{2}.  \label{9.2}
\end{eqnarray}
It provides
\begin{eqnarray}
n_{\mu}u_{\nu} T^{\mu\nu}(x)= \frac{1}{2} \left (\frac{\partial
\phi}{\partial \tau}\right )^{2}+ \frac{1}{2}
\frac{1}{\tau^{2}}\left (\frac{\partial \phi}{\partial \eta}\right
)^{2} +  \frac{1}{2}\left (\frac{\partial \phi}{\partial
\textbf{r}_{T}}\right )^{2}+ \frac{1}{2}m^{2}\phi^{2}. \label{10}
\end{eqnarray}
It follows from eq. (\ref{9.2}) that conjugate momentum with respect
to $\tau$ is $\Pi^{[\tau]}=\frac{\partial \phi}{\partial \tau}$.
Then one can notice that
\begin{eqnarray}
\int_{\sigma_{\nu}} d\sigma n_{\nu}(x)u_{\mu}(x)T^{\mu \nu}(x)=
H^{[\tau]}, \label{10.1}
\end{eqnarray}
and as a result
\begin{eqnarray}
\rho=Z^{-1} \exp\left(-\beta H^{[\tau]}\right ), \label{10.2}
\end{eqnarray}
where $H^{[\tau]}$ is the Hamiltonian that generates translation in
the time-like direction with respect to $\tau$. One sees that such a
Hamiltonian (and the corresponding metrics, see eq. (\ref{5})) is
explicitly $\tau$-dependent. Despite the metrics under the
coordinate transformation (\ref{3}), (\ref{4}) is reduced to the
flat Minkowski metrics, it is not the case for $H^{[\tau]}$ which is
not reduced to $H^{[t]}$.

It follows immediately from eq. (\ref{8}) that  $\phi (x)$ satisfies
to the Klein-Gordon equation of motion
\begin{eqnarray}
(\square - m^{2})\phi (x) =0,  \label{11}
\end{eqnarray}
where $\square = - \partial_{\mu}\partial^{\mu}$ is the d'Alembert
operator associated with the Minkowski spacetime. It is well known
that solution of eq. (\ref{11}) can be written as
\begin{eqnarray}
\phi(x)  =
\int\frac{d^{3}p}{\sqrt{2\omega_{p}}}\frac{1}{(2\pi)^{3/2}}\left
(e^{-i\omega_{p}t+i \textbf{p} \textbf{r}} a (\textbf{p})+
e^{i\omega_{p}t-i \textbf{p} \textbf{r}} a^{\dag}(\textbf{p})\right
), \label{11.1}
\end{eqnarray}
where $\omega_{p}=\sqrt{\textbf{p}^2 + m^{2}}$. The conjugated field
momentum at the hypersurface $t=\mbox{const}$ is
$\Pi^{[t]}=\frac{\partial \phi}{\partial t}$.  The quantization
prescription at such a hypersurface,
\begin{eqnarray}
[\phi(x), \Pi^{[t]}(x')] = i \delta^{(3)}(\textbf{r}-\textbf{r}'),
\label{12}
\end{eqnarray}
means that  functions $a^{\dag}(\textbf{p})$ and $a(\textbf{p})$
become creation and annihilation operators, respectively, which
satisfy the following canonical commutation relations:
\begin{eqnarray}
[a(\textbf{p}), a^{\dag}(\textbf{p}')] =
\delta^{(3)}(\textbf{p}-\textbf{p}'), \label{13}
\end{eqnarray}
and $[a(\textbf{p}), a(\textbf{p}'),]=[a^{\dag}(\textbf{p}),
a^{\dag}(\textbf{p}')]=0$.

It is worth noting that plane-wave mode representation  used in eq.
(\ref{11.1}) has special meaning in our approach  because
corresponding particles are the observed ones, and therefore we are
interested in expectation values for appropriate products of
operators $a^{\dag}(\textbf{p})$ and $a(\textbf{p})$. With this aim,
it is convenient to find representation of the canonical commutation
relations (with corresponding mode functions)  that (at least,
approximately) diagonalizes $H^{[\tau]}$. Because finally we are
interested in expectation values for $a^{\dag}(\textbf{p})$ and
$a(\textbf{p})$, such a representation should be explicitly related
with the plane mode representation (\ref{11.1}). In a most simple
way it can be done if we just rewrite eq. (\ref{11.1}) for
appropriate mode functions.

With this aim  let us first introduce momentum rapidity $\theta$ and
transverse mass $m_{T}$ instead of one-particle energy $\omega_{p}$
and longitudinal momentum $p_{z}$, then
\begin{eqnarray}
\omega_{p}= m_{T} \cosh \theta, \label{14} \\
p_{z}= m_{T} \sinh  \theta ,  \label{15} \\
m_{T}= \sqrt{\textbf{p}_{T}^2 + m^{2}}, \label{16}
\end{eqnarray}
where $\textbf{p}_{T}=(p_{x},p_{y})$ is the transverse momentum.
Then we introduce new operators $\alpha(\textbf{p}_{T},\theta)$,
$\alpha^{\dag}(\textbf{p}_{T},\theta)$ as
\begin{eqnarray}
\alpha(\textbf{p}_{T},\theta) = (m_{T}\cosh \theta)^{1/2}a(\textbf{p}), \label{17} \\
\alpha^{\dag}(\textbf{p}_{T},\theta)=(m_{T}\cosh
\theta)^{1/2}a^{\dag}(\textbf{p}) , \label{18}
\end{eqnarray}
with the commutation relation
\begin{eqnarray}
[\alpha(\textbf{p}_{T},\theta), \alpha^{\dag}(\textbf{p}'_{T},\theta
')] =\delta (\theta - \theta ')
\delta^{(2)}(\textbf{p}_{T}-\textbf{p}_{T}'). \label{19}
\end{eqnarray}
The solution (\ref{11.1}) of the Klein-Gordon equation  (\ref{11})
can be  expressed in terms of these new operators as follows
\begin{eqnarray}
\phi(x)  = \int\frac{d \theta d^{2}p_{T}}{\sqrt{2}(2\pi)^{3/2}}
[\exp(-im_{T}\cosh \theta t+ im_{T}\sinh \theta z + i \textbf{p}_{T}
\textbf{r}_{T}) \alpha(\textbf{p}_{T},\theta) + \nonumber
\\ \exp(im_{T}\cosh \theta t -  im_{T}\sinh \theta z - i \textbf{p}_{T}
\textbf{r}_{T}) \alpha^{\dag}(\textbf{p}_{T},\theta)]. \label{20}
\end{eqnarray}
The next step is to introduce operators $b(\textbf{p}_{T},\mu)$,
$b^{\dag}(\textbf{p}_{T},\mu)$ that are related to operators
$\alpha(\textbf{p}_{T},\theta)$,
$\alpha^{\dag}(\textbf{p}_{T},\theta)$ through the following
formulas
\begin{eqnarray}
\alpha(\textbf{p}_{T},\theta) = \frac{1}{\sqrt{2 \pi}}\int_{-\infty}^{+\infty} e^{i\mu \theta}
b(\textbf{p}_{T},\mu) d \mu , \label{21} \\
\alpha^{\dag}(\textbf{p}_{T},\theta) = \frac{1}{\sqrt{2
\pi}}\int_{-\infty}^{+\infty} e^{- i\mu \theta}
b^{\dag}(\textbf{p}_{T},\mu) d \mu  . \label{22}
\end{eqnarray}
One can see that
\begin{eqnarray}
[b(\textbf{p}_{T},\mu), b^{\dag}(\textbf{p}'_{T},\mu ')] =\delta
(\mu- \mu ') \delta^{(2)}(\textbf{p}_{T}-\textbf{p}_{T}'),
\label{23}
\end{eqnarray}
with all other commutators vanishing. Using the definitions
(\ref{21}), (\ref{22}) one can rewrite eq. (\ref{20}) in the form
\begin{eqnarray}
\phi(x)  = \int\frac{d \theta d^{2}p_{T}d\mu}{\sqrt{2}(2\pi)^{2}}
[\exp(-im_{T}\tau \cosh (\theta - \eta) +  i \textbf{p}_{T}
\textbf{r}_{T} + i \mu \theta ) b(\textbf{p}_{T},\mu)+ \nonumber \\
\exp (im_{T}\tau \cosh (\theta - \eta) -  i \textbf{p}_{T}
\textbf{r}_{T} - i \mu \theta) b^{\dag}(\textbf{p}_{T},\mu) ].
\label{24}
\end{eqnarray}
Performing change of integration variables in eq. (\ref{24}),
$\theta = \vartheta + \eta$, we get
\begin{eqnarray}
\phi(x)  = \int_{-\infty}^{+\infty}\frac{d^{2}p_{T}d\mu
}{4\pi\sqrt{2}}[- i \exp(\mu \pi /2 +i\mu\eta
+i\textbf{p}_{T}\textbf{r}_{T})H^{(2)}_{i\mu}(m_{T}\tau)b(\textbf{p}_{T},\mu)+
\nonumber \\
 i\exp(- \mu \pi /2 - i\mu\eta -
i\textbf{p}_{T}\textbf{r}_{T})H^{(1)}_{i\mu}(m_{T}\tau)b^{\dag}(\textbf{p}_{T},\mu)
], \label{25}
\end{eqnarray}
where $H^{(2)}_{i\mu}(m_{T}\tau)$  and $H^{(1)}_{i\mu}(m_{T}\tau)$
are the Hankel functions \cite{Hankel},
\begin{eqnarray}
H^{(2)}_{i\mu}(m_{T}\tau)= - \frac{1}{i \pi}\exp(-\mu\pi
/2)\int_{-\infty}^{+\infty}d\vartheta \exp(-im_{T} \tau \cosh\vartheta +i\mu \vartheta), \label{26} \\
H^{(1)}_{i\mu}(m_{T}\tau) = \frac{1}{i \pi}\exp(\mu\pi
/2)\int_{-\infty}^{+\infty}d\vartheta \exp( im_{T} \tau
\cosh\vartheta  -i\mu \vartheta). \label{27}
\end{eqnarray}
Using eqs. (\ref{23}), (\ref{25})  and accounting for properties of
the Hankel functions, one can see  that such a representation
realizes the quantization procedure on the hypersurface $\tau =
\mbox{const}$:
\begin{eqnarray}
[\phi(x), \tau \frac{\partial \phi}{\partial \tau}(x')] =i \delta
(\eta- \eta ') \delta^{(2)}(\textbf{r}_{T}-\textbf{r}_{T}').
\label{28}
\end{eqnarray}
The  mode representation (\ref{25}) in the future light cone is well
known, see e.g. refs.
\cite{Milne,Unruh-rev-1,Unruh-rev-2,Unruh-rev-3}, and the vacuum
state with respect to operators $b$ and  $b^{\dag}$ is exactly the
usual Minkowski vacuum.

Substituting (\ref{25}) into (\ref{10}) and performing  integrations
over space-time variables  we bring $H^{[\tau]}$, see eqs.
(\ref{7}), (\ref{10}), and  (\ref{10.1}),  to a form
\begin{eqnarray}
H^{[\tau]}= \frac{\tau \pi}{8}\int_{-\infty}^{+\infty}d^{2}p_{T}d\mu
[G_{(2,1)}(b^{\dag}(\textbf{p}_{T},\mu)b(\textbf{p}_{T},\mu)+b(\textbf{p}_{T},\mu)b^{\dag}(\textbf{p}_{T},\mu))-
 \nonumber \\ e^{\mu
\pi}G_{(2,2)} b(\textbf{p}_{T},\mu)b(-\textbf{p}_{T},- \mu)- e^{-
\mu \pi}G_{(1,1)}
b^{\dag}(\textbf{p}_{T},\mu)b^{\dag}(-\textbf{p}_{T},- \mu)],
\label{29}
\end{eqnarray}
where we introduced notations
\begin{eqnarray}
G_{(l,n)}=[\partial_{\tau}H^{(l)}_{i\mu}(m_{T}\tau)][\partial_{\tau}H^{(n)}_{i\mu}(m_{T}\tau)]+
(\frac{\mu^{2}}{\tau^{2}}+m_{T}^{2})H^{(l)}_{i\mu}(m_{T}\tau)H^{(n)}_{i\mu}(m_{T}\tau).
\label{30}
\end{eqnarray}
Equations (\ref{10.2}), (\ref{29}), and (\ref{30}) provide  the
basis for a diagonalization procedure and  particle momentum
spectrum calculations in the next Section.

\section{Ground-state condensate and its effects on particle momentum spectra}

We begin by noting  that freeze-out of pion momentum spectra in
high-multiplicity $p+p$ collisions happens when $\tau \gtrsim 1$ fm
(and when  $\beta \simeq 1/m$ where $m$ is pion mass).   Then, to
simplify matters we can utilize appropriate approximations to the
Hankel functions in eq. (\ref{30}) and then perform approximate
diagonalization of $H^{[\tau]}$, see eq. (\ref{29}). Namely, let us
assume that $m_{T}\tau \gg 1$. Then, applying  the saddle-point
approximation to the Hankel functions we get
\begin{eqnarray}
H^{(2)}_{i\mu}(m_{T}\tau)\approx  \left (\frac{2}{\pi m_{T}\tau}
\right )^{1/2} \left ( 1+ \frac{\mu^{2}}{m_{T}^{2}\tau^{2}}\right
)^{-1/4} \times \nonumber \\ \exp(i\pi/4-\mu \pi/2-im_{T}\tau
\sqrt{1+\mu^{2}/(m_{T}\tau)^{2}}+i\mu \vartheta_{\sigma}),
\label{31}\\
 H^{(1)}_{i\mu}(m_{T}\tau)\approx  \left (\frac{2}{\pi
m_{T}\tau} \right )^{1/2} \left ( 1+
\frac{\mu^{2}}{m_{T}^{2}\tau^{2}}\right )^{-1/4} \times \nonumber \\
\exp(-i\pi/4+\mu \pi/2+im_{T}\tau
\sqrt{1+\mu^{2}/(m_{T}\tau)^{2}}-i\mu \vartheta_{\sigma}).
\label{32}
\end{eqnarray}
where a value of the saddle-point $\vartheta_{\sigma}$ is defined by
the equation
\begin{eqnarray}
\frac{d}{d\vartheta}[-im_{T}\tau \cosh \vartheta + i \mu
\vartheta]=0, \label{33}
\end{eqnarray}
and is given by
\begin{eqnarray}
\sinh \vartheta_{\sigma}=\frac{\mu}{m_{T}\tau}. \label{34}
\end{eqnarray}
Then,  after some simple but long calculations we get in the leading
order in $(m_{T}\tau)^{-1}$
\begin{eqnarray}
H^{[\tau]}\approx\int_{-\infty}^{+\infty}d^{2}p_{T}d\mu[(m_{T} \cosh
\vartheta_{\sigma}) c^{\dag}(\textbf{p}_{T},\mu)c(\textbf{p}_{T},\mu) + \nonumber \\
\frac{1}{4\tau}(c(\textbf{p}_{T},\mu)c(-\textbf{p}_{T},-
\mu)+c^{\dag}(\textbf{p}_{T},\mu)c^{\dag}(-\textbf{p}_{T},- \mu))],
\label{35}
\end{eqnarray}
where we omitted a constant term  taking into account that such a
term is canceled in the expression for the statistical operator
(\ref{10.2}). Also, we  included phase factors in the
correspondingly re-defined annihilation  and creation operators,
\begin{eqnarray}
c(\textbf{p}_{T},\mu)=b(\textbf{p}_{T},\mu)\exp\left(-\frac{i}{8\tau
m_{T}\cosh \vartheta_{\sigma}}-\frac{i}{2}\tau m_{T}\cosh
\vartheta_{\sigma}+\frac{i}{2}\mu\vartheta_{\sigma}\right), \label{36} \\
c^{\dag}(\textbf{p}_{T},\mu)= b^{\dag}(\textbf{p}_{T},\mu)
\exp\left(\frac{i}{8\tau m_{T}\cosh
\vartheta_{\sigma}}+\frac{i}{2}\tau m_{T}\cosh
\vartheta_{\sigma}-\frac{i}{2}\mu\vartheta_{\sigma}\right).
\label{37}
\end{eqnarray}
One can see that $H^{[\tau]}$ contains combinations of operators
such as $c^{\dag}(\textbf{p}_{T},\mu)c^{\dag}(-\textbf{p}_{T},- \mu)$
and $c(\textbf{p}_{T},\mu)c(-\textbf{p}_{T},- \mu)$ which correspond
to the creation and destruction of two particles with zero total
momentum, respectively, due to the expansion. It is interesting to
note  similarity of the expression (\ref{35}) to the corresponding
expressions which describe  evolution of a scalar field in an
expanding Universe, see e.g. refs. \cite{universe-1,universe-2}.

The Hamiltonian (\ref{35}) is rather easy to diagonalize by means of
the canonical  Bogolyubov transformations with real coefficients.
The diagonalization is obtained by defining a new set of creation
and destruction operators $\xi^{\dag}$ and $\xi$:
\begin{eqnarray}
\xi(\textbf{p}_{T},\mu)=
\frac{c(\textbf{p}_{T},\mu)-A(\textbf{p}_{T},\mu)c^{\dag}(-\textbf{p}_{T},-\mu)}
{\sqrt{1-A^{2}(\textbf{p}_{T},\mu)}} , \label{38} \\
\xi^{\dag}(\textbf{p}_{T},\mu)=
\frac{c^{\dag}(\textbf{p}_{T},\mu)-A(\textbf{p}_{T},\mu)c(-\textbf{p}_{T},-\mu)}
{\sqrt{1-A^{2}(\textbf{p}_{T},\mu)}}, \label{39}
\end{eqnarray}
with canonical commutation relations
\begin{eqnarray}
[\xi(\textbf{p}_{T},\mu),\xi^{\dag}(\textbf{p}'_{T},\mu')]=
\delta(\mu - \mu') \delta^{(2)}(\textbf{p}_{T}-\textbf{p}'_{T}),
\label{40}
\end{eqnarray}
and
$[\xi^{\dag}(\textbf{p}_{T},\mu),\xi^{\dag}(\textbf{p}'_{T},\mu')]$=$[\xi(\textbf{p}_{T},\mu),\xi(\textbf{p}'_{T},\mu')]=0$.
Then according to eqs. (\ref{38}), (\ref{39})
\begin{eqnarray}
c(\textbf{p}_{T},\mu)=
\frac{\xi(\textbf{p}_{T},\mu)+A(\textbf{p}_{T},\mu)\xi^{\dag}(-\textbf{p}_{T},-\mu)}
{\sqrt{1-A^{2}(\textbf{p}_{T},\mu)}} , \label{41} \\
c^{\dag}(\textbf{p}_{T},\mu)=
\frac{\xi^{\dag}(\textbf{p}_{T},\mu)+A(\textbf{p}_{T},\mu)\xi(-\textbf{p}_{T},-\mu)}
{\sqrt{1-A^{2}(\textbf{p}_{T},\mu)}} . \label{42}
\end{eqnarray}
Substituting (\ref{41}) and (\ref{42}) into eq. (\ref{35}) allows us
to diagonalize $H^{[\tau]}$ in operators $\xi^{\dag}$ and $\xi$.
Such a diagonalization   implies   that $A(\textbf{p}_{T},\mu)$ is a
solution of the quadratic equation.   Choosing the solution which
tends to zero when $ m_{T}\tau$ tends to infinity, we get
\begin{eqnarray}
A(\textbf{p}_{T},\mu)=- \frac{1}{4\tau m_{T}\cosh
\vartheta_{\sigma}} .\label{43}
\end{eqnarray}
Under this transformation the Hamiltonian $H^{[\tau]}$  in the
leading order in $(m_{T} \tau)^{-1}$  takes the form
\begin{eqnarray}
H^{[\tau]}\approx \int_{-\infty}^{+\infty}d^{2}p_{T}d\mu (m_{T}\cosh
\vartheta_{\sigma})\xi^{\dag}(\textbf{p}_{T},\mu)\xi(\textbf{p}_{T},\mu).
\label{44}
\end{eqnarray}
A direct consequence of eqs. (\ref{41}) and (\ref{42}) is that the
notion of a vacuum is not unique for ``$c$'' and ``$\xi$''
particles. Namely, the vacuum of ``$c$'' particles coincides with
the ordinary Minkowski vacuum defined with respect to the plane-wave
modes. On the other hand,  vacuum with respect to ``$\xi$''
particles is well-known in the context of quantum
optics\footnote{Such a state also appears in the context of the
Bogolyubov's microscopical theory of superfluidity, see e.g. ref.
\cite{super}}  two-mode squeezed state (see e.g. refs.
\cite{state-1,state-2}) of ``$c$'' particles. Therefore, ground
state of the Hamiltonian $H^{[\tau]}$ is in leading order in $(\tau
m_{T})^{-1}$ a highly entangled state (condensate) of correlated
pairs of $c^{\dag}(\textbf{p}_{T},\mu)$ and
$c^{\dag}(-\textbf{p}_{T},-\mu)$ quanta with zero total momentum.

We now come back to our main objective: use the quasi equilibrium
statistical operator $\rho$, see eq. (\ref{10.2}),  to calculate the
two-boson Bose-Einstein correlation function.  For such calculations
it is convenient to use the method proposed in refs.
\cite{Wick-1-1,Wick-1-2} (see also refs. \cite{Bog,Groot}). First,
let us introduce operator
\begin{eqnarray}
\xi^{\dag}(\textbf{p}_{T},\mu,\beta)=e^{-\beta
H^{[\tau]}}\xi^{\dag}(\textbf{p}_{T},\mu) e^{\beta H^{[\tau]}}.
\label{45}
\end{eqnarray}
It follows from
\begin{eqnarray}
\frac{\partial \xi^{\dag}(\textbf{p}_{T},\mu,\beta)}{\partial
\beta}=[\xi^{\dag}(\textbf{p}_{T},\mu,\beta), H^{[\tau]}]= -
m_{T}\cosh\vartheta_{\sigma
}\xi^{\dag}(\textbf{p}_{T},\mu,\beta)\label{46}
\end{eqnarray}
that
\begin{eqnarray}
 \xi^{\dag}(\textbf{p}_{T},\mu,\beta) = e^{-\beta
m_{T}\cosh\vartheta_{\sigma }}\xi^{\dag}(\textbf{p}_{T},\mu).
\label{47}
\end{eqnarray}
Using  trace invariance under the cyclic permutation of an operator,
we get
\begin{eqnarray}
Tr[ e^{-\beta H^{[\tau]}}
\xi^{\dag}(\textbf{p}_{T1},\mu_{1})\xi(\textbf{p}_{T2},\mu_{2})]=
Tr[ \xi(\textbf{p}_{T2},\mu_{2})e^{-\beta H^{[\tau]}}
\xi^{\dag}(\textbf{p}_{T1},\mu_{1})]=  \nonumber \\  Tr[
\xi(\textbf{p}_{T2},\mu_{2})e^{-\beta H^{[\tau]}}
\xi^{\dag}(\textbf{p}_{T1},\mu_{1})e^{\beta H^{[\tau]}}e^{-\beta
H^{[\tau]}}] = \nonumber \\  Tr[e^{-\beta H^{[\tau]}}
\xi(\textbf{p}_{T2},\mu_{2})e^{-\beta H^{[\tau]}}
\xi^{\dag}(\textbf{p}_{T1},\mu_{1})e^{\beta H^{[\tau]}}].
\label{48.1}
\end{eqnarray}
Using this equation together with  eqs. (\ref{2}), (\ref{10.2}) and
(\ref{45}), one has
\begin{eqnarray}
\langle
\xi^{\dag}(\textbf{p}_{T1},\mu_{1})\xi(\textbf{p}_{T2},\mu_{2})\rangle
= \langle
\xi(\textbf{p}_{T2},\mu_{2})\xi^{\dag}(\textbf{p}_{T1},\mu_{1},\beta)
\rangle .\label{48}
\end{eqnarray}
Substituting (\ref{47}) into  (\ref{48}) and using commutation
relation (\ref{40}) we get
\begin{eqnarray}
\langle
\xi^{\dag}(\textbf{p}_{T1},\mu_{1})\xi(\textbf{p}_{T2},\mu_{2})\rangle
= e^{-\beta m_{T1}\cosh\vartheta_{\sigma 1}}\times \nonumber
\\ (\langle
\xi^{\dag}(\textbf{p}_{T1},\mu_{1})\xi(\textbf{p}_{T2},\mu_{2})\rangle
+ \delta(\mu_{1} - \mu_{1})
\delta^{(2)}(\textbf{p}_{T1}-\textbf{p}_{T2}) ). \label{49}
\end{eqnarray}
It then follows from eq. (\ref{49}) that
\begin{eqnarray}
\langle
\xi^{\dag}(\textbf{p}_{T1},\mu_{1})\xi(\textbf{p}_{T2},\mu_{2})\rangle
= \frac{\delta(\mu_{1} - \mu_{1})
\delta^{(2)}(\textbf{p}_{T1}-\textbf{p}_{T2})}{e^{\beta
m_{T1}\cosh\vartheta_{\sigma 1}}-1}. \label{50}
\end{eqnarray}
Notice that $\langle \xi^{\dag}\xi^{\dag}\rangle = \langle
\xi\xi\rangle = \langle \xi^{\dag}\rangle = \langle \xi \rangle= 0
$. Other $n-$point operator expectation values can be calculated in
a similar way.

Now, let us utilize eqs. (\ref{17}), (\ref{18}), (\ref{21}),
(\ref{22}), (\ref{36}), (\ref{37}), (\ref{41}), and (\ref{42}) to
relate operators $a$ and $a^{\dag}$ with $\xi$ and $\xi^{\dag}$. We
obtain
\begin{eqnarray}
a(\textbf{p}) = \frac{1}{\sqrt{2 \pi m_{T}\cosh
\theta}}\int_{-\infty}^{+\infty}d\mu \left
(\frac{\xi(\textbf{p}_{T},\mu)+A\xi^{\dag}(-\textbf{p}_{T},-\mu)}{\sqrt{1-A^{2}}}\right)
\times  \nonumber
\\  \exp\left(i\mu \theta +\frac{i}{8\tau m_{T}\cosh
\vartheta_{\sigma}}+\frac{i}{2}\tau m_{T}\cosh
\vartheta_{\sigma}-\frac{i}{2}\mu\vartheta_{\sigma}\right),\label{51}
\end{eqnarray}
and
\begin{eqnarray} a^{\dag}(\textbf{p}) = \frac{1}{\sqrt{2 \pi m_{T}\cosh
\theta}}\int_{-\infty}^{+\infty}d\mu \left
(\frac{\xi^{\dag}(\textbf{p}_{T},\mu)+A\xi(-\textbf{p}_{T},-\mu)}{\sqrt{1-A^{2}}}\right
) \times \nonumber
\\\exp\left(- i\mu \theta
-\frac{i}{8\tau m_{T}\cosh \vartheta_{\sigma}}- \frac{i}{2}\tau
m_{T}\cosh \vartheta_{\sigma}+
\frac{i}{2}\mu\vartheta_{\sigma}\right). \label{52}
\end{eqnarray}
Let us recall that the creation and annihilation operators for the
pions are vectors in isotopic spin space,
$\textbf{a}^{\dag}(\textbf{p})=(a_{1}^{\dag}(\textbf{p}),a_{2}^{\dag}(\textbf{p}),a_{3}^{\dag}(\textbf{p}))$,
$\textbf{a}(\textbf{p})=(a_{1}(\textbf{p}),a_{2}(\textbf{p}),a_{3}(\textbf{p}))$,
and  $[a^{\dag}_{i},a_{j}]=0$ if $i\neq j$. Then the annihilation
and creation operators for the charged (say, $\pi^{+}$) pions are
\begin{eqnarray}
a_{+}(\textbf{p}) = \frac{1}{\sqrt{2 }}(a_{1}(\textbf{p})+ i
a_{2}(\textbf{p}) ), \label{53}\\
 a^{\dag}_{+}(\textbf{p}) = \frac{1}{\sqrt{2 }}(a^{\dag}_{1}(\textbf{p})- i a^{\dag}_{2}(\textbf{p}) ), \label{54}
\end{eqnarray}
where $a_{i}$, $a_{i}^{\dag}$ are related with $\xi_{i}$,
$\xi_{i}^{\dag}$ by means of eqs. (\ref{51}), (\ref{52}), here $i$
is $1$ or $2$ and $[\xi^{\dag}_{i},\xi_{j}]=0$ if $i\neq j$.
Utilization of eqs. (\ref{53}), (\ref{54}) and isotopic symmetry
yield  $\langle
a^{\dag}_{+}(\textbf{p}_{1})a_{+}(\textbf{p}_{2})\rangle =\langle
a^{\dag}_{1}(\textbf{p}_{1})a_{1}(\textbf{p}_{2})\rangle$ and
$\langle a_{+}(\textbf{p}_{1})a_{+}(\textbf{p}_{2})\rangle=\langle
a^{\dag}_{+}(\textbf{p}_{1})a^{\dag}_{+}(\textbf{p}_{2})\rangle=0$.
Then using eqs.  (\ref{50}) (\ref{51}), (\ref{52}) we find at
$m_{T}\tau \gg 1$
\begin{eqnarray}
\langle a^{\dag}_{+}(\textbf{p}_{1})a_{+}(\textbf{p}_{2})\rangle
=\frac{\delta^{(2)}(\textbf{p}_{T1}-\textbf{p}_{T2})}{2\pi
\sqrt{\omega_{p1}\omega_{p2}}}\int_{-\infty}^{+\infty}
\frac{e^{-i\mu(\theta_{1}-\theta_{2})}}{1-A^{2}}
\left(\frac{1+A^{2}}{e^{\beta m_{T1}\cosh\vartheta_{\sigma
}}-1}+A^{2}\right ) d\mu . \label{56}
\end{eqnarray}
Also, after some basic but lengthy operator algebra one can see that
the two-particle momentum spectrum,  $\langle
a^{\dag}_{+}(\textbf{p}_{1})a^{\dag}_{+}(\textbf{p}_{2})a_{+}(\textbf{p}_{1})a_{+}(\textbf{p}_{2})\rangle$,
can be written as
\begin{eqnarray}
\langle
a^{\dag}_{+}(\textbf{p}_{1})a^{\dag}_{+}(\textbf{p}_{2})a_{+}(\textbf{p}_{1})a_{+}(\textbf{p}_{2})\rangle
= \langle
a^{\dag}_{+}(\textbf{p}_{1})a_{+}(\textbf{p}_{1})\rangle\langle
a^{\dag}_{+}(\textbf{p}_{2})a_{+}(\textbf{p}_{2})\rangle + \nonumber \\
\langle
a^{\dag}_{+}(\textbf{p}_{1})a_{+}(\textbf{p}_{2})\rangle\langle
a^{\dag}_{+}(\textbf{p}_{2})a_{+}(\textbf{p}_{1})\rangle .
\label{57}
\end{eqnarray}
The above expressions, (\ref{56}) and (\ref{57}), allows us to
estimate the Bose-Einstein  correlation function  for identical
charged bosons (e.g. $\pi^{+}$)  which is defined as
\begin{eqnarray}
C(\textbf{p},\textbf{q})= \frac{\langle
a^{\dag}_{+}(\textbf{p}_{1})a^{\dag}_{+}(\textbf{p}_{2})a_{+}(\textbf{p}_{1})a_{+}(\textbf{p}_{2})\rangle}{\langle
a^{\dag}_{+}(\textbf{p}_{1})a_{+}(\textbf{p}_{1})\rangle\langle
a^{\dag}_{+}(\textbf{p}_{2})a_{+}(\textbf{p}_{2})\rangle},
\label{57.1}
\end{eqnarray}
where $\textbf{p}=(\textbf{p}_{1}+\textbf{p}_{2})/2$,
$\textbf{q}=\textbf{p}_{2}-\textbf{p}_{1}$.

First, notice that our findings imply that one-particle momentum
spectrum can be approximated by  sum of   local equilibrium ideal
gas  and ground-state condensate contributions, where the
ground-state condensate formally corresponds to particle momentum
spectrum at zero temperature. Indeed, one can rewrite eq. (\ref{56})
to the form
\begin{eqnarray} \langle
a^{\dag}_{+}(\textbf{p}_{1})a_{+}(\textbf{p}_{2})\rangle = \langle
a^{\dag}_{+}(\textbf{p}_{1})a_{+}(\textbf{p}_{2})\rangle_{l.eq.}+\langle
a^{\dag}_{+}(\textbf{p}_{1})a_{+}(\textbf{p}_{2})\rangle_{cond},
\label{58}
\end{eqnarray}
where  the last term does not vanish when temperature goes to zero.
Then taking into account that $A\ll 1$ and retaining in each term
only the leading power of $(m_{T}\tau)^{-1}$, one has
\begin{eqnarray}
\langle
a^{\dag}_{+}(\textbf{p}_{1})a_{+}(\textbf{p}_{2})\rangle_{l.eq.}
=\frac{R_{T}^{2}}{(2\pi)^{3}
\sqrt{\omega_{p1}\omega_{p2}}}\int_{-\infty}^{+\infty}d\mu
e^{-i\mu(\theta_{1}-\theta_{2})} \frac{1}{e^{\beta
m_{T1}\cosh\vartheta_{\sigma }}-1} ,  \label{59}\\
\langle
a^{\dag}_{+}(\textbf{p}_{1})a_{+}(\textbf{p}_{2})\rangle_{cond}
=\frac{R_{T}^{2}}{(2\pi)^{3}
\sqrt{\omega_{p1}\omega_{p2}}}\int_{-\infty}^{+\infty}d\mu
e^{-i\mu(\theta_{1}-\theta_{2})}A^{2} . \label{60}
\end{eqnarray}
Here we substitute $\delta^{(2)}(\textbf{p}_{T1}-\textbf{p}_{T2})$
by $(2\pi)^{-2}R_{T}^{2}$ at $\textbf{p}_{T1} = \textbf{p}_{T2}$.
Then, taking into account  eq. (\ref{34}), one can change
integration variable, $\mu = (m_{T}\tau)\sinh(\eta-\theta)$, and get
\begin{eqnarray}
\langle a^{\dag}_{+}(\textbf{p})a_{+}(\textbf{p})\rangle_{l.eq.} =
\frac{R_{T}^{2}}{(2\pi)^{3} \omega_{p}}\int_{-\infty}^{+\infty}d\eta
m_{T}\tau\cosh (\eta -\theta )\frac{1}{e^{\beta m_{T}\cosh
(\eta-\theta)}-1}=\nonumber \\
\frac{1}{(2\pi)^{3} \omega_{p}}\int_{\sigma^{\mu}}d\sigma
u^{\mu}p_{\mu}\frac{1}{e^{\beta u^{\mu}p_{\mu}}-1}, \label{61}
\end{eqnarray}
where $\frac{1}{(2\pi)^{3}}\frac{1}{e^{\beta u^{\mu}p_{\mu}}-1}$
corresponds to the Bose-Einstein  local equilibrium distribution
function of the ideal gas.

It is instructive  to write approximate expressions for one- and
two-particle momentum spectra for $\beta m_{T}\gg 1$ and assuming
$\textbf{p}_{T1} = \textbf{p}_{T2} $ (then $\textbf{q}_{T} =
\textbf{0} $). In such an approximation (\ref{59}) is given by
\begin{eqnarray}
\langle
a^{\dag}_{+}(\textbf{p}_{1})a_{+}(\textbf{p}_{2})\rangle_{l.eq.}
\approx n_{l.eq.}(p)\exp\left (-\frac{m_{T}^{2}\tau^{2}}{2 \beta
m_{T}}(\theta_{1}-\theta_{2})^{2}\right ), \label{62}
\end{eqnarray}
and calculation of (\ref{60}) results in
\begin{eqnarray}
\langle
a^{\dag}_{+}(\textbf{p}_{1})a_{+}(\textbf{p}_{2})\rangle_{cond}\approx
n_{cond}(p) \exp(-m_{T}\tau |\theta_{1}-\theta_{2}|).\label{63}
\end{eqnarray}
Here the one-particle momentum spectra $n_{l.eq.}(p)$ and
$n_{cond}(p)$ are approximately given by
\begin{eqnarray}
n_{l.eq.}(p)=  \frac{R_{T}^{2}}{(2\pi)^{3} \omega_{p}}\tau m_{T}
\sqrt{\frac{2\pi }{\beta m_{T}}}\exp(-\beta m_{T}),\label{64}\\
n_{cond}(p)=\frac{R_{T}^{2}}{(2\pi)^{3}
\omega_{p}}\frac{\pi}{16m_{T}\tau} . \label{65}
\end{eqnarray}
Taking into account that $q_{L}=  p_{2z}-p_{1z}=m_{T}(\sinh
\theta_{2} - \sinh \theta_{1})\approx (m_{T} \cosh \theta )
(\theta_{2} - \theta_{1})$, $\theta = (\theta_{1}+\theta_{2})/2$,
one can rewrite eqs. (\ref{62}) and (\ref{63})  in terms of the more
customary particle momentum difference $q_{L}$,
\begin{eqnarray}
\langle
a^{\dag}_{+}(\textbf{p}_{1})a_{+}(\textbf{p}_{2})\rangle_{l.eq.}
\approx n_{l.eq.}(p_{T})\exp\left (-\frac{R^{2}_{l.eq.}}{2 }q_{L}^{2}\right ), \label{66} \\
\langle
a^{\dag}_{+}(\textbf{p}_{1})a_{+}(\textbf{p}_{2})\rangle_{cond}\approx
n_{cond}(p_{T})\exp\left (-\frac{R_{cond}}{2}|q_{L}|\right ),
\label{67}
\end{eqnarray}
where
\begin{eqnarray}
R_{l.eq.}=\frac{\tau}{\sqrt{\beta m_{T}}\cosh\theta}\label{68}
\end{eqnarray}
is well-known approximate expression for the longitudinal radius
\cite{radii-1,radii-2,radii-3} (see also refs.
\cite{Sin-2-1,Sin-2-2,Sin-2-3,Sin-2-4,Sinyukov-1,Sinyukov-1-1,Sinyukov-2}),
and
\begin{eqnarray}
R_{cond}=\frac{2\tau}{\cosh\theta}\label{69}
\end{eqnarray}
is the longitudinal scale of the ground-state condensate.

The above results allows us to write the correlation function
(\ref{57.1}) as
\begin{eqnarray}
C(\textbf{p}, 0, 0, q_{L})=  1+ \left
(\sqrt{\lambda_{l.eq.}}\exp\left (-\frac{R^{2}_{l.eq.}}{2
}q_{L}^{2}\right )+\sqrt{\lambda_{cond}}\exp\left
(-\frac{R_{cond}}{2}|q_{L}|\right ) \right )^{2}, \label{70}
\end{eqnarray}
where
\begin{eqnarray}
\sqrt{\lambda_{l.eq.}}+\sqrt{\lambda_{cond}} =1, \label{71}
\end{eqnarray}
and
\begin{eqnarray}
\lambda_{l.eq.}=\left
(\frac{n_{l.eq.}}{n_{l.eq.}+n_{cond}}\right )^{2}, \label{72} \\
\lambda_{cond}=\left (\frac{n_{cond}}{n_{l.eq.}+n_{cond}}\right
)^{2}. \label{73}
\end{eqnarray}
It follows from eqs. (\ref{70}) and (\ref{71}) that intercept of the
Bose-Einstein  correlation function  for identical charged pions at
zero relative momentum is equal to $2$,
$C(\textbf{p},\textbf{0})=2$.

It is instructive to compare expression (\ref{70})  with the fitted
form of the correlation function of two identical charged pions,
which in each $\textbf{p}$-bin looks like
\begin{eqnarray}
C_{exp}(\textbf{p},\textbf{q}) =  1+ \lambda_{\textbf{p}}
F_{\textbf{p}}(\textbf{q}),
 \label{74}
\end{eqnarray}
with the  function $F_{\textbf{p}}(\textbf{q})$ depending on the
shape of the boson source, $F_{\textbf{p}}(\textbf{0})=1$ and
$F_{\textbf{p}}(\textbf{q})\rightarrow 0 $ for
$|\textbf{q}|\rightarrow \infty$,   the latter  condition follows
from  normalization of the correlation function that is applied by
experimentalists: $C_{exp}(\textbf{p},\textbf{q}) \rightarrow
 1$ for $|\textbf{q}| \rightarrow \infty$.
One can see that $C(\textbf{p},\textbf{q})$, see eqs. (\ref{56}),
(\ref{57}), and (\ref{57.1}), satisfies the proper normalization
condition.  The phenomenological "chaoticity" parameter $0 \leq
\lambda_{\textbf{p}}\leq 1$ describes the correlation strength. An
estimate of the size of the emission region, in the form of the HBT
radii, is then extracted from a fit of eq. (\ref{74}) in wide enough
$\textbf{q}$-interval  to the measured in each $\textbf{p}$-bin
two-particle correlations with, typically, Gaussian or exponential
from of the function $F_{\textbf{p}}(\textbf{q})$.

Comparing eqs. (\ref{70}) and (\ref{74}), we see that, strictly
speaking, we can not identify $\lambda$-parameters and radii in eq.
(\ref{70}) with the measured ones.  Nevertheless, just to have an
idea on the magnitude of the effect of the ground-state condensate
on the two-particle correlation function, one can rewrite eq.
(\ref{70}) to the form
\begin{eqnarray}
C(\textbf{p}, 0, 0, q_{L})= 1+ \Lambda(p_{T},q_{L})
e^{-R^{2}_{l.eq.}q_{L}^{2}} , \label{75}
\end{eqnarray}
where by definition
\begin{eqnarray}
\Lambda(p_{T},q_{L}) =\left (1 - \sqrt{\lambda_{cond}}\left ( 1-
e^{-\frac{R_{cond}}{2}|q_{L}|+ \frac{R^{2}_{l.eq.}}{2 }q_{L}^{2}}
\right )\right )^{2} . \label{76}
\end{eqnarray}
Identification of $\Lambda(p_{T},q_{L})$ with phenomenological
parameter $\lambda_{\textbf{p}}$ can give us rough estimate of the
latter in the region of the correlation peak: $|q_{L}|\sim
1/R_{l.eq.}$, $|q_{T}| = 0$. Namely, it follows from eq. (\ref{76})
that in such a region $\Lambda(p_{T},q_{L})<1$ and decreases when
transverse momentum of a pion pair increases, and that there is the
correlation between the size of the emission region and
$\Lambda(p_{T},q_{L})$: for smaller radius, $R_{l.eq.}$, one can see
that $\Lambda(p_{T},q_{L})$ is also smaller. Namely, when transverse
momentum increases, then $\lambda_{cond}$ increases, see eqs.
(\ref{64}), (\ref{65}) and  (\ref{73}). Also, $R_{l.eq.}< R_{cond}$,
and $R_{l.eq.}$ decreases when transverse momentum increases while
$R_{cond}$ is constant, see eqs. (\ref{68}) and (\ref{69}), as a
consequence $\Lambda(p_{T},q_{L})$ decreases for $|q_{L}|\sim
1/R_{l.eq.}$ when $p_{T}$ increases. Such a behavior seems to be at
least qualitatively consistent  with behavior of the
phenomenological $\lambda_{\textbf{p}}$-parameter derived from
experimental parametrization (\ref{74}) \cite{Atlas,CMS}. For very
high transverse  momenta the ground-state condensate contribution to
one-particle momentum spectrum dominates, and
$\lambda_{l.eq.}\rightarrow 0$, $\lambda_{cond}\rightarrow 1$ in the
two-particle correlation function (\ref{70}) when $p_{T}\rightarrow
\infty$, see eqs. (\ref{64}), (\ref{65}), (\ref{72}) and (\ref{73}).

Finally, let us address the question how our results are changed if
one uses next-to-leading order terms in $(m_{T}\tau)^{-1}$. First at
all, note that  the ground-state condensate appears if the ground
state (at zero temperature) of quasi equilibrium statistical
operator (\ref{1}) does not coincide with the ordinary vacuum state
in Minkowski spacetime, in the model considered it is the case
because the ground state with respect to $H^{[\tau]}$ is not the
same as the ground state (vacuum) with respect to the ordinary
Hamiltonian. While we have calculated  the  momentum spectra of
over-condensate excitations  as well as condensate particles only in
the leading order of perturbative expansion in $(m_{T}\tau)^{-1}$,
one can conclude that accounting for next orders in
$(m_{T}\tau)^{-1}$ does not change general qualitative features of
our results. Namely, the two-component parametrization of the
two-particle momentum correlation function, see eq. (\ref{70}),
remains unchanged, as well as decrease of
$\frac{n_{l.eq.}}{n_{cond}}$ and $\frac{R_{l.eq.}}{R_{cond}}$ when
transverse momentum increases. Then, in particular,  one can
conclude that estimation of $\Lambda(p_{T},q_{L})$ behavior with
$p_{T}$ remains correct.

\section{Conclusions}

Motivated by the recent experimental observations of the suppression
of two-particle Bose-Einstein momentum correlations in the
high-multiplicity $p+p$ collisions at the LHC
\cite{Alice-1,Atlas,CMS}, we relate these observations to the
ground-state condensate contribution to particle momentum spectra.
Our findings demonstrate  that under certain circumstances quasi
equilibrium expansion of quantum fields can lead to the condensate
formation. We considered a model for particles emission, based on
the hypothesis of quasi equilibrium boost-invariant longitudinal
expansion. In the present work, to make the problem tractable, we
used the model of non-interacting scalar quantum field  to describe
the sudden particle momentum spectra freeze-out at the hypersurface
$\tau = \mbox{const}$, and  do not take into account transverse
finite-size effects.  In our opinion the proposed physical picture
reflects in some degree the physics of the quasi equilibrium
evolution and particles emission in the high-multiplicity $p+p$
collisions at the LHC. We demonstrated that the ground-state
condensate can lead to suppression of the two-particle Bose-Einstein
momentum correlations due to the two-scale mechanism of particles
emission.\footnote{Notice that two-scale mechanism of particles
emission  was also proposed in ref. \cite{scale-1} based on
different underlying physical picture, and two-particle
Bose-Einstein correlations in $p+p$ collisions at the LHC  were
fitted by the correspondingly parameterized correlation function
\cite{scale-2}.} This effect should be less important for
relativistic heavy ion collisions which have prolonged
post-hydrodynamical kinetic stage of hadronic rescatterings. To
demonstrate the reliability  of the model proposed, we performed a
simple analytical estimate of the apparent $\lambda$-parameter  in
the two-pion Bose-Einstein correlation function. The main point here
is not a detail comparison with experimental data. To make it
possible one needs to take into account that the particle-emitting
sources produced in high-energy $p+p$ collisions are expanding also
in the transverse direction, particle momentum spectra include
feed-downs from the resonance decays, etc., but such an analysis
goes beyond the scope of the present article. Our main finding is
that if state of a system can be approximated by the quasi
equilibrium statistical operator of the longitudinally
boost-invariant expanding quantum scalar fields, then the
boost-invariant  hydrodynamics is accompanied by the ground-state
condensate. This should be taken into account when the
hydrodynamical approach is applied to analyze high-energy $p+p$
collision experiments.

\begin{acknowledgments}
I am grateful to Yu.M. Sinyukov for discussions.
\end{acknowledgments}

\end{document}